\documentclass[doublecol]{epl2} 
\usepackage[numbers]{natbib}
\usepackage[]{amsmath}
\usepackage[]{hyperref}

\title{A faster scaling in acceleration-sensitive atom interferometers}
\shorttitle{A faster scaling in acceleration-sensitive atom interferometers}

\author{G. D. McDonald\thanks{E-mail: \email{gordon.mcdonald@anu.edu.au}}, C. C. N. Kuhn, S. Bennetts, J. E. Debs, K. S. Hardman, J. D. Close \and N. P. Robins}
\shortauthor{G. D. McDonald, \etal}

\institute{Quantum Sensors Lab, Department of Quantum Science, Research School of Physics and Engineering, Australian National University, Canberra, 0200, Australia.  Website: \href{http://atomlaser.anu.edu.au/}{http://atomlaser.anu.edu.au/}
}

\abstract{
Atom interferometers have been used to measure acceleration with at best a $T^2$ scaling in sensitivity as the interferometer time $T$ is increased. This limits the sensitivity to acceleration which is theoretically achievable by these configurations for a given frequency of acceleration. We predict and experimentally measure the acceleration-sensitive phase shift of a large-momentum-transfer atom interferometer based upon Bloch oscillations. Using this novel interferometric scheme we demonstrate an improved scaling of sensitivity which will scale as $T^3$. This enhanced scaling will allow an increase in achievable sensitivity for any given frequency of an oscillatory acceleration signal, which will be of particular use for inertial and navigational sensors, and proposed gravitational wave detectors. A straight forward extension should allow a $T^4$ scaling in acceleration sensitivity.
}

\begin{document}

\maketitle

Cold-atom interferometers which measure accelerations have many applications at the cutting edge of technology and fundamental physics. These range from inertial sensing and navigation~\citep{RobinsPhysicsReports,Geiger:2011,Peters:1999aa,BestAtomicGravimeter}, geodesy and climate monitoring though atomic equivalents of the GRACE project \citep{GraceClimate}, to measurements of the fine structure constant~\citep{Alpha2011,ANDP:ANDP201300044,MassInSeconds}, gravitational wave detection~\citep{BramGrav},  and other tests of general relativity~\citep{MullerGravityTest}. 
These devices measure the acceleration $\mathbf{a}$ which would be experienced by a freely accelerating test mass $m$, by comparing the accumulated phase of atoms travelling along each of the two space-time trajectories which make the interferometer. As the two trajectories are coincident at the beginning and end of the interferometer, they enclose an area in space-time, defined by $\mathbf{\mathcal{A}}=\int \Delta \mathbf{x}\, dt$ (see Fig. 1). In a separate paper \citep{SpaceTimeAreaArxiv} we use the Feynman path-integral formalism to show that the relative phase of such an interferometer can be written as $\Phi=\frac{m}{\hbar}\mathbf{\mathcal{A}\cdot a}$, under quite general time-symmetry conditions~\citep{footnote1}. Therefore, the phase sensitivity to acceleration is given by $\frac{d\Phi}{da}=\frac{m}{\hbar}\mathcal{A}$.

\begin{figure}[!htbp]
\centering{}
  \includegraphics[width=1\columnwidth]{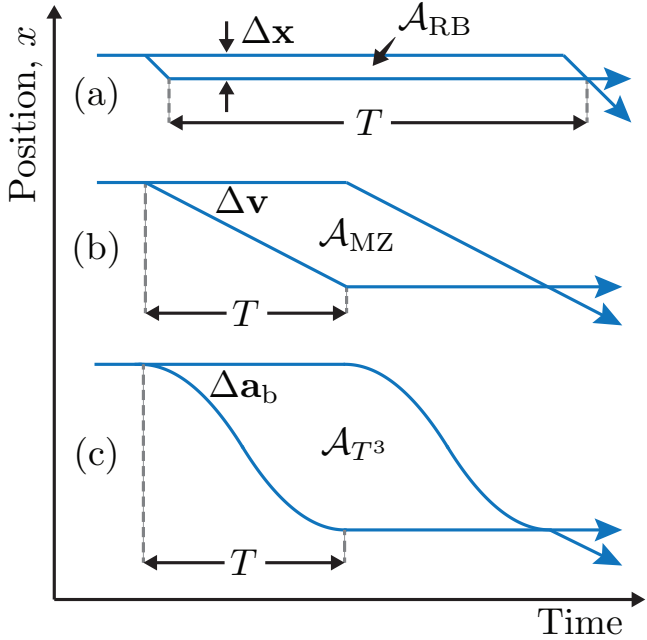}
 \caption{The space time area $\mathbf{\mathcal{A}}=\int \Delta \mathbf{x}\, dt$ is illustrated for several acceleration-sensitive interferometer configurations. For clarity, the common inertial acceleration $\mathbf{a}$ is zero in this diagram.  (a) In a Ramsey-Bord\'e configuration, the two particle trajectories are separated by a constant displacement $\Delta \mathbf{x}$ for the duration of the interferometer.  (b) A Mach-Zehnder configuration separates the two trajectories by a constant velocity difference $\Delta \mathbf{v}$. Halfway through the interferometer, this velocity difference reverses sign.  (c) This paper introduces the configuration in which the two trajectories are separated by a constant acceleration $\Delta \mathbf{a}_{b}$.}
\label{Space-time Diagram}
\end{figure}

Two acceleration sensitive configurations are in common use. The Ramsey-Bord\'e (RB) configuration  \citep{Borde1989,gBloch,CladeLMT,HMullerBlochBraggBloch,SchmiedmayerBECMZI,CladeBounce,Muller24hk} [shown in Fig. 1 (a)] is constructed from two trajectories which stay a constant displacement $\Delta \mathbf{x}$ apart for the duration $T$ of the interferometer, neglecting the quick accelerations required to initially separate and finally recombine the two paths. It therefore has a space-time area $\mathbf{\mathcal{A}}_{{RB}}=\Delta \mathbf{x} T$  and so its acceleration sensitivity scales linearly with $T$. 
The Mach-Zehnder (MZ) configuration \citep{FirstAtomInterferometer,Muller24hk,DoubleDiffraction,OurGravimeter,DebsBECgrav,80hkPRA,SelfWaveguide,ResolveRedshift} [Fig. 1 (b)] is constructed from two paths that have a constant velocity difference $\Delta \mathbf{v}$, which is reversed after the interferometer time $T$. The MZ space-time area is given by $\mathbf{\mathcal{A}}_{{MZ}}=\Delta \mathbf{v}T^2$ and so its sensitivity to acceleration will scale with $T^2$. 

In this letter we present the logical extension: an interferometer configuration in which the two paths are separated by a constant acceleration $\Delta \mathbf{a}_{b}$ (up to a sign change) with respect to one another [as shown in Fig.~1~(c)]. In this configuration, the space-time area is given by $\mathbf{\mathcal{A}}_{T^3}=\frac{\Delta \mathbf{a}_{b}}{4}T^3$ and so the sensitivity to the common, external acceleration $\mathbf{a}$ will scale as $T^3$. Our experimental implementation of this technique uses Bloch oscillations to affect the continuous acceleration $\mathbf{a}_{b}$. This Continuous-Acceleration Bloch (CAB) technique combines this new geometry with an MZ configuration to achieve a space time area of $\mathbf{\mathcal{A}}_{{CAB}}=\mathbf{\mathcal{A}}_{{MZ}}+\mathbf{\mathcal{A}}_{T^3}=\Delta \mathbf{v}T^2+\frac{\Delta \mathbf{a}_{b}}{4}T^3$. Therefore, our technique will allow greater sensitivity than an MZ configuration for any given interferometer time $T$ and scales as a higher power of $T$, thereby increasing the phase sensitivity to accelerations at any given frequency of accelerations to be measured. Of all the applications for atom-interferometric accelerometers mentioned, inertial sensing, navigation and gravitational wave detection are the three which will immediately benefit from this increased sensitivity at a given frequency. We will now experimentally demonstrate this faster scaling in phase sensitivity to acceleration as compared with a standard MZ configuration. 
Extension of the technique to higher scalings in $T$ are also discussed.

Atom interferometers utilising large momentum separation require a source cloud of atoms with a narrow momentum width to maintain interferometer visibility~\citep{SzigetiBragg}. 
Our technique for generating such a cold atom source is detailed in Ref.~\citep{80hkPRA}.
 Briefly, we form a $2\times10^6$ atom Bose-Einstein condensate of $^{87}$Rb in a cross-beam dipole trap [see Fig. 2 (a)], before turning off the cross beam to load the condensate into the remaining dipole trapping beam, the optical waveguide. 
 An optical delta-kick cooling sequence is employed followed by velocity selection Bragg pulse to create a cold cloud with a longitudinal momentum width below $0.05\hbar \mathbf{k}$. 

 The optical lattice used to apply the interferometry sequence to the atoms is described in Ref.~\citep{SelfWaveguide}. 
We have up to 50mW in each of two counter-propagating laser beams aligned collinear with the waveguide, detuned 105 GHz to the blue from the $\left|F=1\right> \rightarrow\left|F'=2\right>$ transition of the $D_2$ line in $^{87}\text{Rb}$, which keeps the number of spontaneous emissions below 1\% of our total atom number during our interferometric sequence. 
Arbitrary, independent control of the frequency detuning and amplitude of each beam is achieved using a direct digital synthesiser.

\section{MZ Interferometry sequence} 
\begin{figure*}
\begin{center}
\includegraphics[width=2\columnwidth]{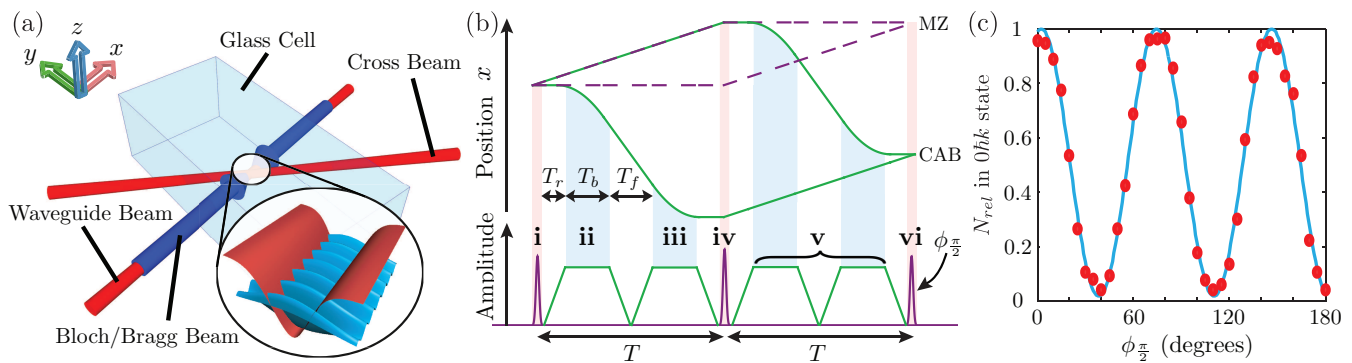}
 \caption{(a) A BEC is formed in a cross beam dipole trap (red), before it is released into the horizontal waveguide. The interferometer is formed by optical pulses from the Bragg/Bloch beam (blue). Inset: The potential (not to scale) experienced by the atoms when both the waveguide and the optical lattice are on. (b) Top: The space-time trajectory of each arm of the interferometer (total time $2T$) is shown for both a $10{\hbar}\mathbf{k}$ MZ configuration (purple dashed lines), and the CAB sequence (green solid lines) for the same time $T$. The vertical axis represents position along the horizontal waveguide in the experiment. Bottom: Our pulse sequence for the CAB interferometer. $10{\hbar}\mathbf{k}$ Bragg diffraction pulses (purple gaussians) are the beam splitter ($\frac{\pi}{2}$, outside) and mirror ($\pi$, centre) pulses. $2n_{b}{\hbar}\mathbf{k}$ Bloch lattice accelerations present in the CAB sequence (green trapezoids) act only upon one arm of the interferometer at a time. (c) The relative population in one of the output states of an MZ with $T=0.2$ms is measured as the phase of the final beamsplitter pulse $\phi_\frac{\pi}{2}$ is scanned. Each realisation of the experiment is shown as a red circle, and a sinusoidal fit to the data of the form $N_{rel}=\frac{\mathcal{V}}{2}\cos(\Phi+n\phi_{\frac{\pi}{2}})+c$ (shown in blue) is used to extract the interferometric phase $\Phi$, which is the acceleration signal.}
\label{Config}
\end{center}
\end{figure*}

A standard MZ interferometer sequence is constructed as follows, each part of which is labeled with roman numerals corresponding to its depiction in Fig~\ref{Config}~(b). 
First, a cold cloud of Rubidium-87 atoms with a narrow momentum width is loaded into a horizontal optical waveguide, which effectively confines them to move freely in only the longitudinal direction ($x$ in Fig.~\ref{Config}). 
 \textbf{i.} An optical standing wave formed from two counter-propagating beams each of wavevector $\pm \mathbf{k}$ is then used as a matter-wave diffraction grating, beam-splitting the atoms along two trajectories which are separated in momenta by the Bragg condition, $\Delta \mathbf{p}_x=2n\hbar \mathbf{k}$, where $n$ is the diffraction order. \textbf{iv.} 
After a time $T$, another Bragg-diffraction pulse is applied which diffracts the stationary state to $2n\hbar \mathbf{k}$ at the same time as diffracting the $2n\hbar \mathbf{k}$ state to be stationary. This effectively swaps the momenta of each state, so it is the atom-optical equivalent of a mirror. 
\textbf{vi.}~When a total time $2T$ has elapsed and both trajectories are overlapping once again, a final Bragg-diffraction beam-splitter pulse is applied which recombines the two momentum states. 
This final optical standing wave can have an arbitrary phase offset $\phi_\frac{\pi}{2}$ from the initial optical standing wave \textbf{i}, which amounts to an $x$-displacement of the peaks and troughs in optical intensity.

\section{CAB interferometry sequence} Our CAB interferometer sequence is built upon this standard MZ configuration. 
\textbf{i.} After the first Bragg-diffraction beam-splitter there are two momentum states in the waveguide, one stationary and one with momentum $\mathbf{p}_x=2n\hbar \mathbf{k}$ in the positive-$x$ direction. 
\textbf{ii.} Over a time $T_{r}$ we selectively load the stationary momentum state into a stationary optical lattice with a potential energy depth of $15E_{r}$, where the photon recoil energy is given by $E_{r}=({\hbar}k)^2/2m$. 
We accelerate the lattice in the negative-$x$ direction via $n_b$ Bloch oscillations \citep{PeikBloch} over a time $T_{b}$, each of which imparts $2\hbar \mathbf{k}$ momentum. 
This amounts to a constant acceleration rate of $\Delta \mathbf{a}_{b}=\frac{2n_{b}\hbar \mathbf{k}}{mT_{b}}$ applied to the state loaded into the lattice. 
The state selectivity is accomplished because the optical lattice is moving fast enough with respect to the  $\mathbf{p}=2n\hbar \mathbf{k}$ momentum state that this state experiences just the time-averaged lattice potential, which imparts no acceleration. 
\textbf{iii.} After a negligible time $T_{f}$ we decelerate the state until it is stationary again. 
\textbf{iv.} The Bragg-diffraction mirror pulse is applied a time $T$ after the beginning of the interferometer, which swaps the momenta of each state. 
\textbf{v.} The Bloch lattice acceleration and deceleration sequence is now applied to the other arm of the interferometer. 
\textbf{vi.} The two momentum states are recombined by a final Bragg-diffraction beam-splitter. It should be noted that the CAB sequence is actually a combination of both configuration (b) and (c) depicted in Figure 1.
	
	To obtain an interferometric fringe like that shown in Fig.~\ref{Config}~(c), we allow the two final momentum states to separate before taking an absorption image to count the number of atoms $N_i$ in each final state $i$. We calculate $N_{rel}=\frac{N_1}{N_1+N_2}$ to remove the effect of run-to-run fluctuations in total atom number of $\approx15\%$. As the phase $\phi_\frac{\pi}{2}$ of the final beam-splitter is scanned, a fringe will be obtained of the form $N_{rel}=\frac{\mathcal{V}}{2}\cos(\Phi+n\phi_{\frac{\pi}{2}})+c$ as shown in Fig.~\ref{Config}~(c). The phase offset $\Phi$ of this fringe is our acceleration signal.

 The acceleration dependence of this phase offset is given by the space-time area $\mathcal{A}_{\text{CAB}}=\int \Delta \mathbf{x}\, dt$ of our interferometer,

\begin{align}
      \Phi_{CAB}&=\frac{m}{\hbar}\mathbf{a}\cdot\mathbf{\mathcal{A}}_{CAB}\nonumber\\
       &=2\left(n+n_{b}\cdot \frac{T_{b}+T_{f}}{T}\right)\mathbf{k\cdot a}T^2
	\label{eq:phase}
\end{align}
which reduces to the phase offset of an MZ configuration $\Phi_{MZ}=2n\mathbf{k\cdot a}T^2$ when $n_{b}=0$.

	In order to obtain the $T^3$ scaling, we must look at what happens when the interferometer time $T$ is increased, keeping constant the relative acceleration $\Delta \mathbf{a}_{b}$ which we apply via the Bloch lattice. In this case $n_{b}=T_{b}/\tau$, where the constant $\tau=\frac{2{\hbar}k}{m|\Delta \mathbf{a}_{b}|}$ is the period for one Bloch oscillation. Assuming $T_{r}$ and $T_{f}$ are much smaller than $T_{b}$, then we have $T_{b}\rightarrow\frac{T}{2}$, and Eq.~(\ref{eq:phase}) becomes
	
\begin{align}
	\label{eq:T3phase}
	\Phi_{CAB}&=2n\mathbf{k\cdot a}T^2+\frac{\mathbf{k\cdot a}T^3}{2\tau}\\
	&=\Phi_{MZ}+\frac{\mathbf{k\cdot a}T^3}{2\tau}\nonumber
\end{align}
which explicitly shows the extra $T^3$ scaling in acceleration sensitivity achievable with this configuration.

\begin{figure}[!htp]
\centering{}
  \includegraphics[width=.9\columnwidth]{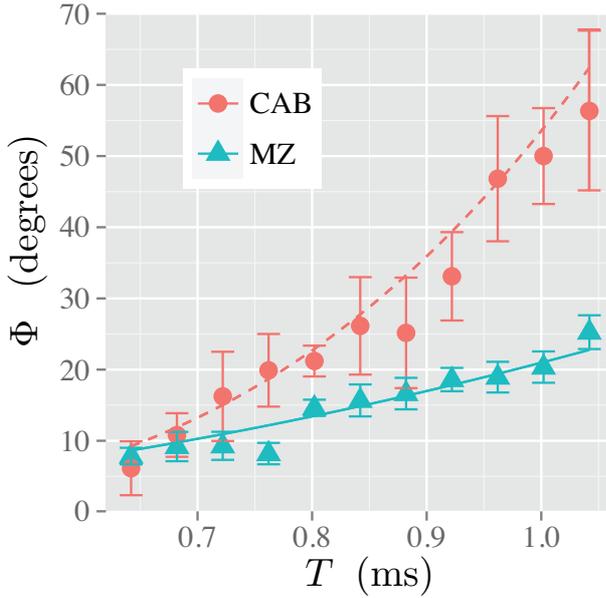}
 \caption{Here we experimentally demonstration the faster sensitivity scaling. The interferometric phase offset $\Phi$ is measured for each fringe as the interferometer time $T$ is increased. The blue solid line is a fit to the MZ phase $\Phi_{MZ}$ (blue triangles) of the form $\Phi_{MZ}=2n\mathbf{k\cdot a}T^2$ to extract the acceleration $a$ along the waveguide (due to a slight tilt) for the day the data was taken. This acceleration is then used to predict $\Phi_{CAB}$ for the CAB sequence by equation (\ref{eq:phase}) (red dashed line). Experimental measurement of $\Phi_{CAB}$ for the CAB sequence (red circles) increases faster than $T^2$, as predicted by Eq. (\ref{eq:phase}). Uncertainties in $\Phi$ extracted from each fringe are one standard deviation confidence intervals.}
\label{MoneyShot}
\end{figure}

We experimentally test Eq.~(\ref{eq:phase}) by first measuring $\Phi_{MZ}$ for a standard MZ configuration with $n=5$, $n_{b}=0$ (blue triangles on Fig.~\ref{MoneyShot}) in order to extract the external acceleration $\mathbf{a}$ due to a tilt in the waveguide. From this we calculate via Eq.~\ref{eq:phase} what $\Phi_{CAB}$ will be for the CAB sequence (red dashed line). We then measure $\Phi_{CAB}$ for $T=0.642$ms, $n_{b}=1$ through to $T=1.042$ms, $n_{b}=21$ (red circles) and our experimental parameters were set such that $T=2n_{b}\tau+0.622$ms with a Bloch oscillation period of $\tau=10\mu$s. The excellent agreement shown in Fig.~\ref{MoneyShot} between the predicted phase for the CAB sequence which was deduced from the MZ measurements, and the measurements of $\Phi_{CAB}$ validate Equation (\ref{eq:phase}) and demonstrate a $T^3$ scaling in phase sensitivity to acceleration.

We now turn to yet higher scalings with respect to $T$.
The maximum adiabatic acceleration rate $\Delta \mathbf{a}_{b}$ using a Bloch lattice increases quadratically with lattice depth \citep{PeikBloch}, and therefore also increases quadratically with available laser power. 
This means that a Bloch-based constant acceleration configuration can achieve a larger space-time area for constant $T$ and a given laser power than an equivalent sequential-Bragg configuration \citep{Kasevich102hk}, in which the momentum transferrable in each Bragg diffraction pulse increases as the square root of the available laser power \citep{SzigetiBragg}. 
However, in the CAB scheme lattice depth and acceleration rate are limited by the instantaneous velocity separation of the two clouds \citep{80hkPRA}. 
This is because if the lattice is too deep it will also bind the other momentum state, and thus our acceleration will no longer be state selective.
Therefore, the best possible use of available laser power would require the application of a constantly increasing relative acceleration $\Delta\frac{d\mathbf{a}_{b}}{dt}$ (known as constant jerk) between the two states, while commensurately increasing the optical lattice depth. 
This would give a a space-time area of $\mathcal{A}_{T^4}=\frac{1}{24}\Delta\frac{d\mathbf{a}_{b}}{dt} T^4$, producing an interferometer with phase $\Phi=\Phi_{MZ}+\frac{m}{\hbar}\mathbf{a}\cdot\mathcal{A}_{T^4}$. 
At the point at which lattice depth is limited by available laser power, the maximum acceleration rate will become constant again, and so the scaling will revert to $T^3$. 
Generalising to arbitrary scalings in $T$ would be possible if an interferometer were developed with a constant $n$-th derivative of displacement (for $n\geq 1$). 
It would have a space time area of 

\begin{align}
	\label{eq:generalisation}
    \mathcal{A}_{T^{n+1}}&=\frac{1}{n!\,2^{n-1}}\left[ \left(\frac{d}{dt}\right)^{n}  \Delta \mathbf{x}\right]T^{n+1}
\end{align}
and therefore a $T^{n+1}$ scaling in acceleration sensitivity. Although there is no practical advantage beyond constant jerk in the present system, it is possible that an analogous scheme will be developed in the future which is not limited by laser power, and could practically take advantage of these higher scalings. For instance, one can envisage positionally dependent trapping potentials (e.g. dipole traps) accelerating each cloud in opposite directions once they are separated in space.

The choice of Bragg~\cite{80hkPRA} and not Raman~\cite{CladeLMT} beamsplitters in this configuration allows both momentum states to be in the same magnetic internal state throughout the interferometer, eliminating a systematic phase shift. The use of a BEC source cloud presents its own systematic phase shift, the density-dependent mean-field shift~\cite{KyleCoherence}. However, this shift can be reduced arbitrarily by lowering the cloud's density via the delta-kick-cooling process~\cite{80hkPRA}, or removed entirely by turning off mean-field interactions through the use of a Feshbach resonance~\cite{SoltionArxiv}. To separate the effects of acceleration due to gravity, the optical potential and the magnetic field gradients possibly present in the experiment, one could use a magnetically insensitive internal state~\cite{SelfWaveguide} or compare the phase shift across different isotopes simultaneously~\cite{DualSpecies}.

In the future, optical-lattice intensity-noise reduction in this system is possible by using multiple overlaid Bloch lattices to address both momentum states separately, and accelerate them in opposite directions at the same time~\citep{HMullerBlochBraggBloch}. 
This will cancel the a.c. Stark shift due to Bloch laser intensity noise, as each arm simultaneously experiences the Bloch lattice. In the present CAB sequence the a.c. Stark shift is only cancelled at a later time in the interferometer, so with this improvement the interferometer will become less sensitive to such fluctuations. However, as each arm of the interferometer only experiences half the laser intensity, this technique will quarter the maximum acceleration rate of each state, due to it quadratic scaling with intensity \citep{PeikBloch}. This results in a halving of the relative acceleration rate $\Delta\mathbf{a}_{b}$, and hence a halving of the signal $\Phi$. Other noise reductions in order to enhance signal-to-noise can be achieved by reducing mechanical vibrations~\cite{80hkPRA}, evacuating the optical path and locking-out optical-lattice-laser frequency fluctuations~\cite{GbThesis}.

In summary we have demonstrated a novel configuration for a cold-atom interferometer in which acceleration sensitivity scales as $T^3$. This CAB configuration is realised using an optical Bloch lattice to subject one arm of the interferometer at a time to an additional constant acceleration. The additional $T^3$ scaling in sensitivity to the external inertial acceleration $\mathbf{a}$ allows this CAB configuration to have increased sensitivity to accelerations measured with a given interferometer time $T$ than what is possible with a Mach-Zehnder configuration. This CAB configuration will therefore be useful in increasing the phase sensitivity to accelerations at any given frequency, without requiring any increase in available laser power. This technique can therefore be immediately applied to greatest effect in navigation and inertial sensors which are currently under development, and in proposed schemes for gravitational wave detection.

\acknowledgments
The authors would like to thank Hannah Keal for her assistance moving electronic equipment out of the lab, and Paul Altin for baking the vacuum system. We gratefully acknowledge the support of the Australian Research Council Discovery program. C.C.N.K. would like to acknowledge financial support from CNPq (Conselho Nacional de Desenvolvimento Cientifico e Tecnologico). J.E.D. would like to acknowledge financial support from the IC postdoctoral fellowship program.


\begin{thebibliography}{10}
\bibitem{footnote1}
  {Consider the interferometer to be centered at time $t=0$, i.e. the first splitting occurs at $t=-T$ and the recombination occurs at $t=T$. If $\mathbf{\alpha}_1(t)$ is the acceleration of cloud 1 and $\mathbf{\alpha}_2(t)$ is the acceleration of cloud 2, the time symmetry required is $\frac{1}{2}\left[\mathbf{\alpha}_1(-t)+\mathbf{\alpha}_2(t)\right]=\mathbf{a}$. All interferometer configurations depicted in Fig. 1 obey this symmetry.}
\end{thebibliography}

\end{document}